\documentclass[conference]{IEEEtran}
\IEEEoverridecommandlockouts
\usepackage{cite}
\usepackage{amsmath,amssymb,amsfonts}
\usepackage{algorithmic}
\usepackage{graphicx}
\usepackage{textcomp}
\usepackage{xcolor}
\def\BibTeX{{\rm B\kern-.05em{\sc i\kern-.025em b}\kern-.08em
    T\kern-.1667em\lower.7ex\hbox{E}\kern-.125emX}}

\begin{document}

\title{Hand gesture recognition using 802.11ad mmWave sensor in the mobile device
}

\author{\IEEEauthorblockN{Yuwei Ren$^1$, Jiuyuan Lu$^1$, Andrian Beletchi$^1$, Yin Huang$^1$, Ilia Karmanov$^2$, Daniel Fontijne$^2$, \\ Chirag Patel$^3$ and Hao Xu$^3$}
\IEEEauthorblockA{
\textit{$^1$QUALCOMM Wireless Communication Technologies (China) Limited, Qualcomm AI Research}\\
\textit{$^2$Qualcomm Technologies Netherlands B.V., Qualcomm AI Research}\\
\textit{$^3$Qualcomm Technologies, Inc., Qualcomm AI Research}\\
$\{$ren, ljiuyuan, abeletch, yinh, ikarmano, dfontijn, cpatel and hxu$\}$@qti.qualcomm.com}
}

\maketitle

\begin{abstract}
We explore the feasibility of AI assisted hand-gesture recognition using 802.11ad 60GHz (mmWave) technology in smartphones. Range-Doppler information (RDI) is obtained by using pulse Doppler radar for gesture recognition. We built a prototype system, where radar sensing and WLAN communication waveform can coexist by time-division duplex (TDD), to demonstrate the real-time hand-gesture inference. It can gather sensing data and predict gestures within 100 milliseconds. First, we build the pipeline for the real-time feature processing, which is robust to occasional frame drops in the data stream. RDI sequence restoration is implemented to handle the frame dropping in the continuous data stream, and also applied to data augmentation. Second, different gestures RDI are analyzed, where finger and hand motions can clearly show distinctive features. Third, five typical gestures (swipe, palm-holding, pull-push, finger-sliding and noise) are experimented with, and a classification framework is explored to segment the different gestures in the continuous gesture sequence with arbitrary inputs. We evaluate our architecture on a large multi-person dataset and report $>$ 95\% accuracy with one CNN + LSTM model. Further, a pure CNN model is developed to fit to on-device implementation, which minimizes the inference latency, power consumption and computation cost. And the accuracy of this CNN model is more than 93\% with only 2.29K parameters.
\end{abstract}
\section{Introduction}
\label{sec:intro}
Mobile computing has shown powerful capability, and far exceeds the desktop application scenarios. The human–computer interaction (HCI) community has seen tremendous interest in the mobile-based solution, especially on the smartphone. In the context of mobile and wearable computing, the main focus is the vision based HCI, which have reached high levels of accuracy in stationary settings, but it requires good light condition. Besides, vision based HCI has privacy concerns.

RF-sensing has been developed to match the ubiquitous interaction in mobile application. 
Different to an appearance-based recognition system (e.g. standard video), radar sensing can work well regardless of light conditions with limited power consumption. Also it is not affected by differences in skin color and other static information and is thus less likely to overfit to certain genders and races. However, the capability of the lower frequency system is fundamentally limited by the narrow bandwidth, small antenna aperture, and large wavelength. Specifically, due to the small antenna aperture, the spatial resolution of these systems is too low to distinguish the RF signals reflected by multiple reflectors. 

Recently, higher frequency band is explored in wireless local area networks (WLAN), e.g., 60GHz in 802.11ad, and in cellular, e.g., 28GHz in 5G. Such mmWave band provides larger bandwidth, compact antenna array and lower power consumption, and brings new opportunities to the RF-based HCI, e.g., one promising contactless phone control. 
In the forthcoming 5G and beyond network, such RF-sensing and wireless communications co-existence, also termed as integrated sensing and communications (ISAC) in \cite{cui2020mutual}, is with huge commercial demands but facing diverse challenges. For example, shared spectrum and hardware for the co-existence is designed to ensure no interference between two technologies or performance degradation in \cite{alloulah2019future}.
This paper will show a feasible way to have WLAN and RF-sensing coexistence in mmWave band.

One mature and popular method is Frequency-Modulated Continuous Wave (FMCW) based waveform in the RF sensing. Its transmitted signals is frequency modulated by a periodical saw-wave function. Generally, the frequency shift from the time delay between Tx and Rx would maintain the stable Doppler and range estimation. But the waveform spreads in the whole time duration, and the corresponding power consumption is relative large. Further, if coexisting with the communication waveform, e.g., OFDM in 60GHz, the continuous frequency modulation would require additional RF chain, which isn't available for the coexisting of sensing and communications. Another option is pulse-based waveform, where only small partial time is occupied in the time duration, e.g., 5\% duty cycle, which largely save the power. Moreover, simple pulse waveform is easily modulated by sharing the most RF components in the communication systems. For example, the commonly used random access procedure is designed with the Pulse-based waveform. 

Another important consideration is the antenna deployment. Generally, in HCI scenario, radar array is placed in the front, where the Tx-Rx beam is with high resolution in space. But for mobile device, e.g., the phone, the antenna space is limited, and it is hard to use the front antenna. Even with the front antenna configuration, the screen would largely attenuate the signal strength in the propagation from the internal to outside. With the constraints, antenna are usually assigned along the sides and the back of the phone. Unfortunately, the sides assignment would degrade the beam spatial angle resolution compared to the front.

Embracing these challenges, we build a smartphone based RF sensing prototype, and propose a deep learning based gesture recognition framework, specifically designed for the recognition of dynamic gestures with millimeter wavelength RF signals. The main innovations include:
\begin{itemize}
  \item A data logging pipeline is proposed where series of parameters are optimized to fix the frame dropping issues.
  \item A noise-based segmentation method is designed to fetch clean gestures segments in the sequence dataset.  
  \item Side-based antenna array deployment is implemented, which align with mobile device size constraints.
  \item Pulse-based waveform is applied, coexisting with the base-band processing in single chip.
  \item Finally, one CNN+LSTM sequence model and one tiny pure CNN model are designed to recognize different gestures, including hands and fingers, with high accuracy and small size.
\end{itemize}

The rest of the paper is organized as follows. 
Section 2 introduces the related work in the RF-based hands gesture recognition. 
In section 3, the data pipeline is built to illustrate the theoretical model of CIR, and the RDI sequence processing.
Section 4 proposes an end-to-end trained stack of deep learning (DL) solution in the real-time operation. 
Finally, we conclude our findings in the paper and describe the future work.

\section{Related works}
\label{sec:Related-works}

Many camera based solutions exist, mainly focusing on 2D RGB images \cite{erol2007vision}. \cite{kim2012digits} uses the wrist-worn cameras, and \cite{song2014air} considers the mobile devices built-in cameras. Recently, fine-grained 3D hand pose estimation is experimented in the real-time depth camera, e.g., in \cite{keskin2012hand,sharp2015accurate}. However, the direct line of sight constraints and the privacy concerns have restricted the wide application for the mobile scenarios. 

One of the non-vision based solutions is to leverage the RF signals to detect the motions. In the low frequency band, usually less than 6GHz, \cite{yeo2016radarcat} provides an overview, where disturbances of GSM signals are used for the sensing in \cite{zhao2014sideswipe}, and other examples include picking-up electromagnetic interference in LCDs \cite{chen2013utrack} or piggybacking onto existing WLAN signals \cite{pu2013whole}. But the limiting spatial resolution in low frequency band degrades the performance.

Industrial, Soli chip was commercialized in Google Pixel4, which accelerates the higher frequency band application for gestures recognition at 60GHz. Soli related work \cite{lien2016soli,wang2016interacting,smith2018gesture,flintoff2018single} focuses on the short range hands gesture recognition. \cite{lien2016soli,wang2016interacting} is from Google to introduce the Soli sensor, and its application for gesture detection. \cite{smith2018gesture} introduces the related application for the human-car interaction. \cite{flintoff2018single} proposes a hyper-adaptive robot hand, combined with the Soli sensor and tactile sensor. Besides, NVIDIA develops the FMCW mono-pulse radar works at 24GHz \cite{molchanov2015short}, and based on the TI sensor, it considers the radar and depth camera to detect the gesture. Intel’s work for the gesture recognition in \cite{cai2019efficient} consider the pure CNN network, by constructing the 3 channels 2D image. OPPO and BUPT consider the long-range hands recognition in \cite{liu2020long}. All of these works are purely based on FMCW waveform, and designed with dedicated chip, which needs to be further optimized for the mobile phone application. 


Pulse-based waveform is also experimented in many works. \cite{arbabian201394} provides one HW implementation with pulse-based transmission. \cite{fhager2019pulsed} might be the only one, based on pulse-based gesture recognition with machine learning. \cite{kim2017hand} only consider the reflected signals as the input, with a simple HW configuration.\cite{khan2017hand} provides one demo usage for gesture recognition in the hands-car scenario.

A challenge of gesture recognition is the dynamic modelling in a motion sequence. Recently, a growing trend toward feature representations learning is based on deep neutral networks. For example, sequence based solution, e.g., Long short-term memory (LSTM) cells, has been shown to capture sequential information, which has been widely used for gesture recognition, including Soli related works \cite{lien2016soli,wang2016interacting,smith2018gesture,flintoff2018single} and others. Besides, \cite{liu2019spectrum} investigate the feasibility of human gesture recognition using the spectra of radar measurement parameters. \cite{will2019human} provides one comprehensive analysis based on the conventional method such as support vector machine (SVM) to distinguish the motion and gestures. \cite{hazra2019short} provide one comprehensive work on the RDI based on the 3D to distinguish the hands gesture.

Besides the gesture recognition, in ISAC configuration, mmWave sensing have been explored in many aspects. For example, \cite{cui2020mutual} jointly maximizes the sum-rate for communication and the mutual information between the target impulse response and target echoes for radar. \cite{barneto2020radio} proposes a radio-based sensing approach utilizing the 5G NR uplink transmit signal and an efficient receiver processing and mapping scheme.

\section{System design and theoretical model}

\subsection{Data processing Pipeline}
\label{sec:pipeline}
\begin{figure}[htb]
    \centering
    \includegraphics[scale=0.25]{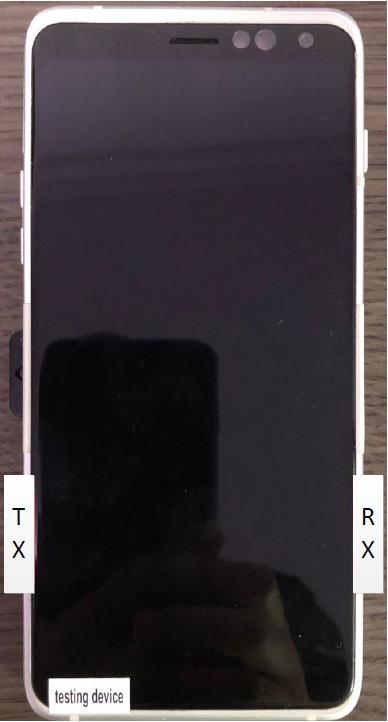}
    \caption{Phone factor implementation equipped with 60GHz and two sides antenna assignment.}
    \label{fig:QRD}
\end{figure}

\begin{figure}[htb]
    \centering
    \includegraphics[scale=0.5]{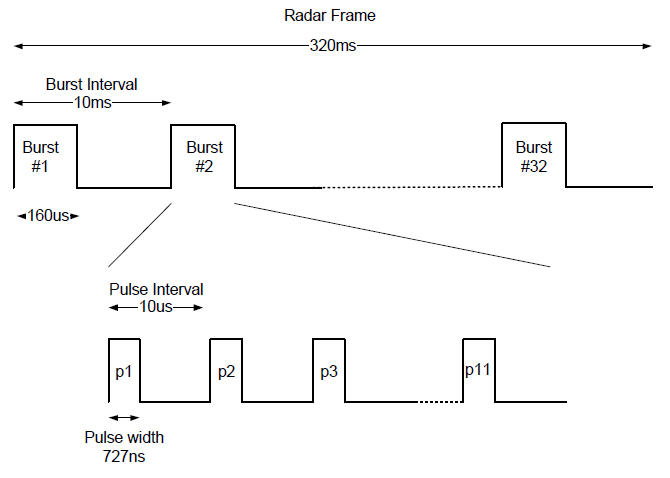}
    \caption{Pulse-based waveform and radar frame configuration.}
    \label{fig:frame_format}
\end{figure}

As shown in Fig.\ref{fig:QRD}, we use a Qualcomm testing device, as our phone factor prototype. 
It is equipped with 802.11ad WLAN 60GHz Chipsets, and assigned antennas at the phone sides. 
The device has 32 elements assembled in a 6×6 layout for both the transmitter (Tx) and the receiver (Rx) and operates with 3.52GHz bandwidth. 
We consider to operate it in radar mode, which is TDD multiplexing with WLAN waveform.
In the radar portion, one radio frame is split into multi-burst interval, and each burst duration is with low cycle duty, shown in Fig.\ref{fig:frame_format}.
The Tx transmits a known pulse sequence in the configured radar burst. 
 
\begin{figure}[htb]
    \centering
    \includegraphics[scale=0.6]{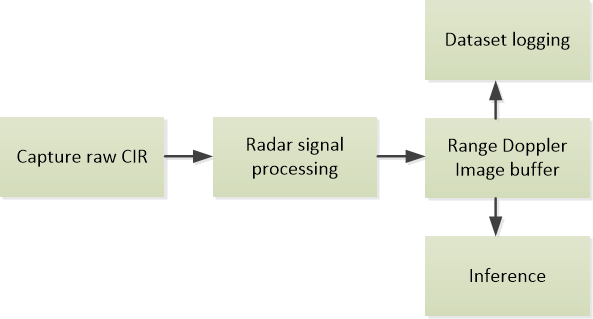}
    \caption{Data processing pipeline}
    \label{fig:functions}
\end{figure}

The Tx transmits a known pulse sequence for channel impulse response (CIR) estimation, and CIR is the source output of the radar device. Refer to the Fig.\ref{fig:functions}, pulse-based solution leverages the CIR to derive the 2D matrix by FFT operation, called as RDI in the Radar signal processing function. The RDI buffer would be sent for either data-set logging module function, or for the DL model for the real-time inference. The RDI image is thought as the input of DL model.

Assume the signals of the electromagnetic waves from Tx antenna $m$ to Rx antenna $n$ are formulated as ${h_{m,n}}\left( t \right)$, where $2R$ is the distance from Tx to Rx after the target reflection, and $c$ is the light speed. To measure frequency, the signal is observed for at least one cycle. To model how $R$ changes with time, assume constant velocity and the distance is expressed as $R = {R_0} + {v_0}t$. Substituting in (\ref{Equ:E1})

\begin{equation}
\label{Equ:E1}
\begin{array}{l}
{h_{m,n}}\left( t \right) = {a_{m,n}}\left( t \right)\exp \left[ { - j2\pi {f_0}\left( {t + \frac{{2R}}{c}} \right)} \right]\\
\quad \quad \;\;\; = {a_{m,n}}\left( t \right)\exp \left[ { - j\left( {2\pi t\left( {{f_0} + {f_0}\frac{{2{v_0}}}{c}} \right) + \frac{{2\pi {f_0}{R_0}}}{c}} \right)} \right]
\end{array}
\end{equation}

\begin{equation}
\label{Equ:E2}
{f_D} =  - {f_0}\frac{{2{v_0}}}{c} =  - \frac{{2{v_0}}}{{{\lambda _0}}}
\end{equation}
where $f_D$ is the Doppler frequency shift. By convention, positive Doppler shift means the target and radar closing.


\subsection{Sequence frame dropping and recovery}
In the experiment, we build one self-contained data-set, where 21 people, with different genders/ages/weights, are invited for the data logging. There are more than 20000 sequences, each with duration of 60 seconds roughly. Although this procedure takes much effort, the data-set is still limited. Besides, there are random frame dropping in the captured RDI sequence. Ideally, sensors capture the data sample-by-sample as one sequence, for example, with index 0,1,2,3…, etc. Hardware constraints (e.g., limited computation capability, and limited power or buffer size designed for mobile phones) may lead to random frame dropping, which destroys the sequence order information. The frame dropping would further reduce the size of the data-set and lead to the loss of useful information. 

\begin{figure}[htb]
    \centering
    \includegraphics[scale=0.4]{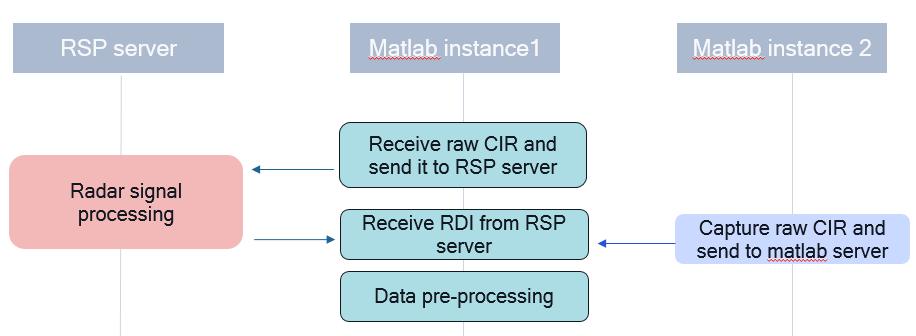}
    \caption{SW parallel pipeline optimization}
    \label{fig:HW-optimization}
\end{figure}

\begin{figure*}[htb]
    \centering
    \includegraphics[scale=0.33]{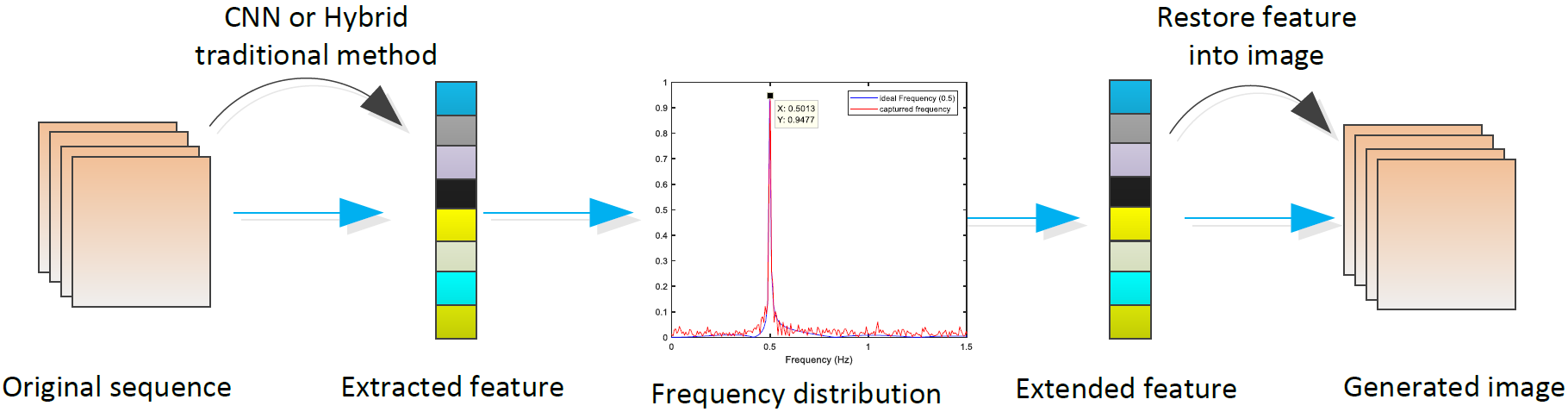}
    \caption{Sequence restoration blocks}
    \label{fig:restoration}
\end{figure*}

Generally, hardware optimization would solve the issue, but this would involve additional large cost. Here, we propose solutions with two parts: one with software (SW) optimization, and another with pattern restoring algorithm in the frequency domain. Fig.\ref{fig:HW-optimization} shows how to allocate the processing function in 3 parallel pipelines, which would improve the frame rate from 4FPS to 7FPS. 
The two instances act as client and sends data stream to server for radar signal processing, which makes client keep gathering data instead of waiting for data processing. 
Further, shown in the Fig.\ref{fig:restoration}, the RDI feature sequence is extracted within the encoder, and each dimension of the feature vector is optimized after the 1-D FFT operation to extend the length of the sequence. And further, with the decoder, the restored sequence is from the extended feature. Here, the encoder and decoder could be one Auto-encoder model \cite{vincent2008extracting}. 

\subsection{Sequence segmentation}

\begin{figure}[htb]
    \centering
    \includegraphics[scale=0.6]{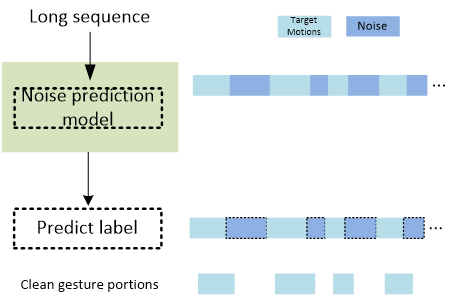}
    \caption{Noise-based effective motions segmentation.}
    \label{fig:segmentation}
\end{figure}

A second challenge presented by the captured RDI sequence is that RDI sequences mix noise portions and target motion portions.  Additionally, the target motion portions are difficult to identify in the mixed sequence, because noise RDI show similar characteristics to those of the target motion, and conventional vision-based solutions are not applicable. Accordingly, in some aspects, noise portions of the input sequence can be easily predicted, identified, and removed to produce a sequence including a less or no noise portions. The resulting sequence may include only the target motion portions (e.g., pure portions). Shown in the Fig.\ref{fig:segmentation}, one binary classification model is used to recognize the noise features in the short-range scenarios. Here, the one-vs-others binary model is based on the pure motions dataset, purely associated to some repeated hands or fingers actions without any noise and body interference. And finally, for any mixed sequence, the pre-trained model could accurately identify and remove the noise portions, and the left is the pure motion sessions. 

All of the above operations are implemented offline, and the complexity would not impact the real-time inference.

\section{DL solutions for the recognition}
\label{sec:DL}
We propose a deep-learning architecture for gesture recognition with high-frequency radar. While our implementation is specific to the WLAN pulse-based implementation, generalizing the approach to other high-frequency RF signals is straightforward.

\begin{figure}[htb]
    \centering
    \includegraphics[scale=1]{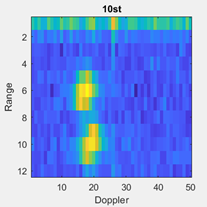}
    \caption{Range Doppler Image with two targets in the range with similar speeds.}
    \label{fig:RDI}
\end{figure}

In the RDI, e.g., Fig.\ref{fig:RDI}, the features are summarized as one channel image, where the detected target with high light pixel, and the X/Y-dimension mapped to the Doppler speed / range index. The shape of the high light pixels set is corresponding to the speed and range variation in the measured frame duration, which is not directly representing any target.

\begin{figure}[htb]
    \centering
    \includegraphics[scale=0.6]{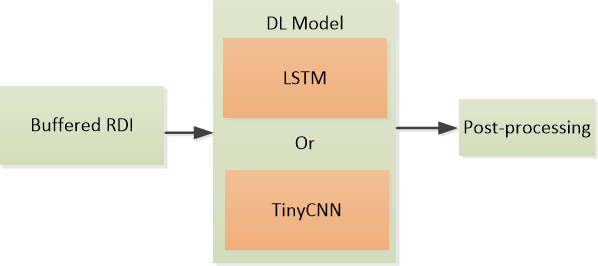}
    \caption{DL structure.}
    \label{fig:ML_mian}
\end{figure}

Generally, continuous RDI sequence would represent the actions. As shown in Fig.\ref{fig:ML_mian}, the buffered RDI are pushed into the DL model, where the corresponding output is classification of the different gestures. And further, one post-processing function is to improve end to end inference accuracy, e.g., label prediction hysteresis-based approach, which could prevent the wrong gesture prediction from transition point between the two different gestures.
For DL model, we explore the CNN + LSTM structure, which is widely utilized in many works,e.g., \cite{wang2016interacting}. And we also propose one tiny CNN model to adapt to the on-device implementation.

In the experiments, we optimize the RDI with size 9*49, which is roughly corresponding to $<$ 40cm range (4cm resolution) and  $<$1m/s speed (2cm/s resolution). Meanwhile, we explore multi buffered sequence length from 5 to 10. 

\subsection{CNN + LSTM}
\label{ssec:LSTM}
\begin{figure}[htb]
    \centering
    \includegraphics[scale=0.25]{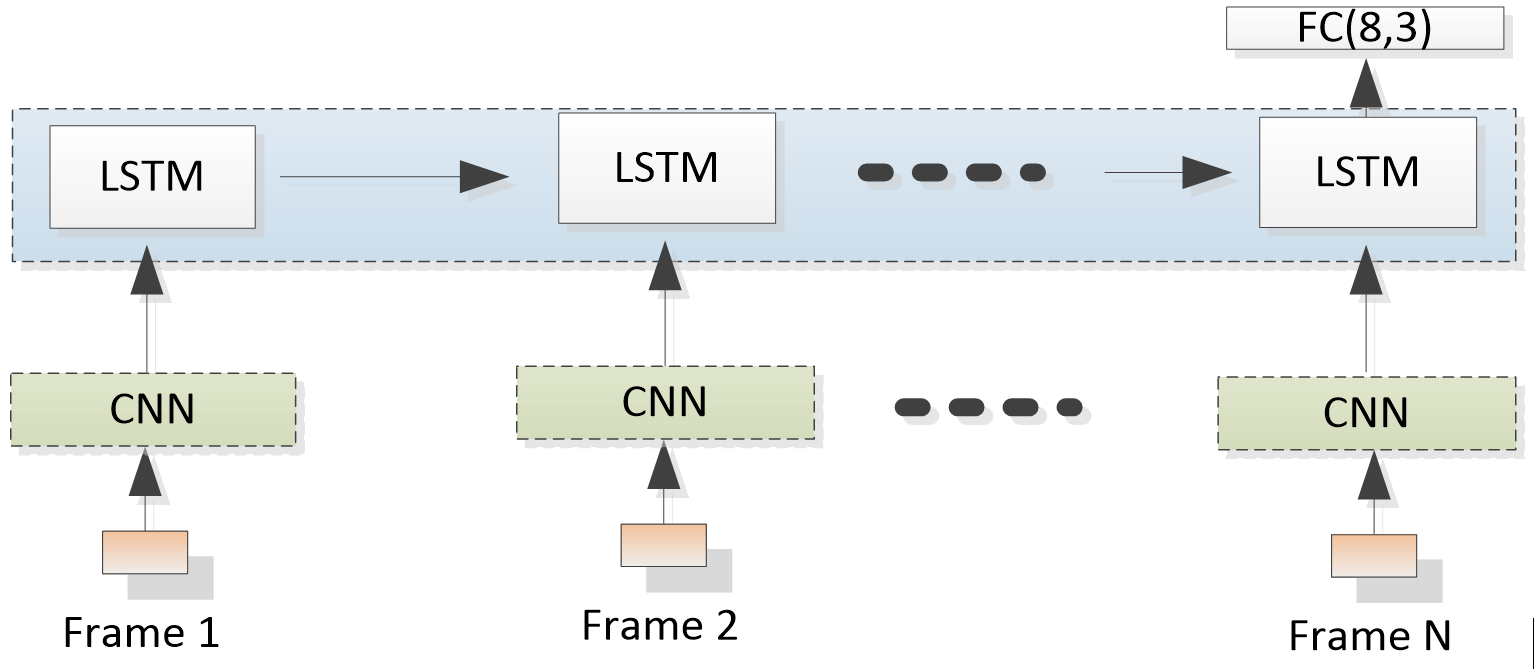}
    \caption{CNN+ LSTM structure with $N$ images input}
    \label{fig:LSTM}
\end{figure}

Conv layer is to extract the 2D feature of RDI in the input sequence, and then, LSTM tracks the temporal features. The detail is shown in the Fig. \ref{fig:LSTM}, where CNN block is to extract the frame feature with two convolution layers: kernel size = $3 \times 3$, dropout ratio = 0.5 and flatten vector output $16 \times 1$.

\subsection{Pure CNN}
\label{ssec:CNN}

\begin{figure}[htb]
    \centering
    \includegraphics[scale=0.3]{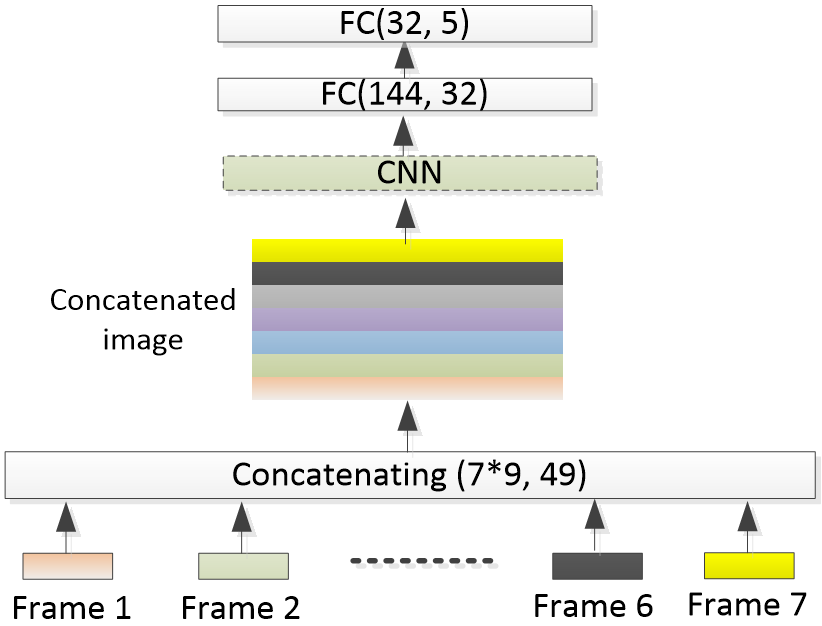}
    \caption{Pure CNN model}
    \label{fig:pure-cnn}
\end{figure}

LSTM shows the good performance with acceptable cost, but sequence model on-device still faces many challenges. Limited memory restricts complicated model application, e.g., the weights size and the operation. Therefore, we propose another pure tiny CNN model, shown in Fig.\ref{fig:pure-cnn}, where 7 frames are combined into one 2D enlarged image with the size 63*49. Here, we build one CNN block with 3 convolution layers to extract the sequence feature, where kernel size = $3 \times 3$, and $Max$ $Pooling$ and $Batch$ $Normalization$ are leveraged to avoid the over-fitting and accelerate the convergence

\subsection{Performance evaluation}
\label{ssec:performance}
\begin{figure}[htb]
    \centering
    \includegraphics[scale=0.45]{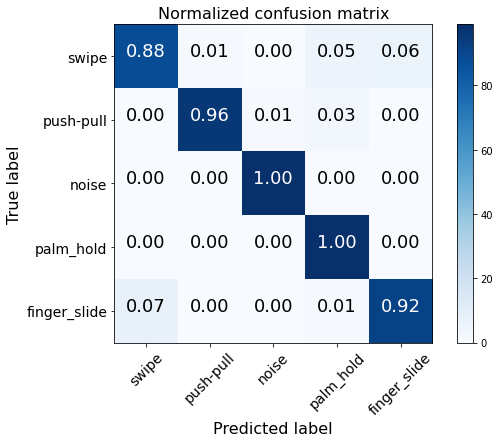}
    \caption{Confusion matrix: CNN + LSTM performance with 5 images input}
    \label{fig:LSTM-per}
\end{figure}

The LSTM model is to explore the feasibility of the gesture recognition in the setting. Simply, we analyze the performance with 5 images in each captured sequence. Validation accuracy is larger than 95\%, shown in Fig. \ref{fig:LSTM-per}. But, in the realistic on-device implementation, LSTM-Cell involves additional specific HW complexity, e.g., forget and input gates, to support the real-time application, which increases additional cost of the implementation in terms of memory and battery.

The LSTM exploration with 5 images has shown that the short sequence has already kept the main feature of one gesture, which roughly match the effective gesture duration, about 0.625 second. For the short memory work, the CNN is another option with the combined inputs.

The pure tiny CNN model is proposed to achieve the low-complexity. Firstly, the input sequence is combined to 2D matrix, frame by frame with one order. One 2D matrix corresponds to one captured sequence. In the experiment, we evaluate different sequence length [3.5,7,10], and the overall validation accuracy is shown in TABLE. \ref{tab:table1}. 
\begin{table}[h!]
  \begin{center}
    \caption{Pure Tiny CNN: Accuracy vs Input sequence length.}
    \label{tab:table1}
    \begin{tabular}{l|c}
      \textbf{Sequence length} & \textbf{Accuracy, \%} \\
      \hline
      3 & 89.96\\
      5 & 93.13\\
      7 & 95.41\\
      10 & 97.15\\
    \end{tabular}
  \end{center}
\end{table}

Although large sequence length improves the accuracy, it involves more computation complexity and latency, shown in Table.\ref{tab:table2}. 
10 images is with more processing time, partially because of the model size $\sim$ 16K,and user experiences some delays of the recognition.
Moreover, the training accuracy approaches 100\%, which might have over-fitting. With the real-time demo, considering the actual quality of the user experience (QoE), such as latency and accuracy, length 7 is the final selection, with larger than 95\% accuracy and only $\sim$ 6K parameters.
\begin{table}[h!]
  \begin{center}
    \caption{Overall performance comparison between the length 5, 7 and 10.}
    \label{tab:table2}
    \begin{tabular}{c|c|c|c}
      \textbf{Sequence length} & 5 & 7 & 10 \\
      \hline
      \textbf{Training accuracy}, \% & $>$95 & $>$98 & $>$99 \\
      \textbf{Validation accuracy}, \% & $>$93 & $>$95 & $>$97 \\
      \textbf{Parameter size} & $\sim$ 2K & $\sim$ 6K & $\sim$ 16K \\
    \end{tabular}
  \end{center}
\end{table}

\begin{figure}[htb]
    \centering
    \includegraphics[scale=0.45]{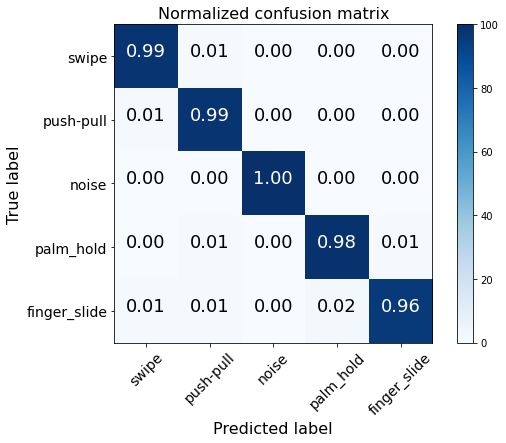}
    \caption{Confusion matrix: pure tiny CNN performance with 7 images input.}
    \label{fig:pure-cnn-per}
\end{figure}

The detail is shown in the confusion matrix, Fig. \ref{fig:pure-cnn-per}. 

\section{Conclusion}
\label{sec:conclusion}
In this paper, we presented a real-time gesture recognition implementation for smartphone applications, where a tiny CNN model was proposed with only 2.29K parameters. 5 different gestures can be detected with the accuracy larger than 93\% in less than 100 millisecond latency. Additionally, we applied sequence restoration to fix the frame dropping issue in the captured sequence, which is also used for the data augmentation. Further, we proposed a noise-based motion segmentation method to get the intentional gesture segments. The current prototype has two-sides antenna array layout, which constraints the beam-forming performance in the front. In future, we will explore beam-forming optimization for 3D hands tracking capabilities.


\bibliographystyle{IEEEbib}
\bibliography{refs}

\end{document}